\documentclass[twocolumn,showpacs,amsmath,amssymb,prb]{revtex4}
\usepackage{graphicx}

\begin{document}

\title{Enhanced vortex heat conductance in mesoscopic
superconductors}
\author{N.~B.~Kopnin $^{(1,2)}$}
\author{A.~S.~Mel'nikov $^{(3)}$}
\author{V.~I.~Pozdnyakova $^{(3)}$}
\author{D.~A.~Ryzhov $^{(3)}$}
\author{I.~A.~Shereshevskii $^{(3)}$}
\author{V.~M.~Vinokur $^{(4)}$}
\affiliation{$^{(1)}$ Low Temperature Laboratory, Helsinki
University of Technology, P.O. Box 2200, FIN-02015 HUT, Finland,\\
$^{(2)}$ L.~D.~Landau Institute for Theoretical Physics RAS,
117940 Moscow, Russia,
 \\
$^{(3)}$ Institute for Physics of Microstructures RAS,  603950,
Nizhny Novgorod, GSP-105,  Russia,
 \\
$^{(4)}$ Argonne National Laboratory, Argonne, Illinois 60439 }

\date{\today}

\begin{abstract}
Electronic heat transport along the flux lines in a long ballistic
mesoscopic superconductor cylinder with a radius of the order of
several coherence lengths is investigated theoretically using both
semiclassical approach and the full quantum-mechanical analysis of
the Bogoliubov--de Gennes equations. The semiclassical approach is
constructed analogously to the Landauer transport theory in
mesoscopic conductors employing the idea that heat is carried by
the quasiparticle modes propagating along the vortex core. We show
that the vortex heat conductance in a mesoscopic sample is
strongly enhanced as compared to its value for a bulk
superconductor; it grows as the cylinder radius decreases. This
unusual behavior results from a strongly increased number of
single-particle transport modes due to giant mesoscopic
oscillations of energy levels, which originate from the interplay
between the Andreev reflection at the vortex core boundary and the
normal reflection at the sample edge. We derive the exact
quantum-mechanical expression for the heat conductance and solve
the Bogoliubov--de Gennes equations numerically. The results of
numerical computations confirm the qualitative Landauer-type
picture and allow us to take into account the partial reflections
of excitations. We analyze the effect of surface imperfections on
the spectrum of core excitations. We show that the giant
oscillations of core levels and thus the essential features of the
heat transport characteristic to ideal mesoscopic samples hold for
a broad class of surface imperfections as well.
\end{abstract}


\maketitle

\section{Introduction}

Low-temperature measurements of the heat transport in the mixed
state of type-II superconductors is a unique tool providing
information about characteristics of the quasiparticle excitations
and superconducting gap structure.  When the external magnetic
field $B$ is applied, thermal transport in the direction of field
is dominated by the quasiparticle waves propagating along the
vortex cores. Thermal conductivity in the mixed state is a
longstanding problem considered in many classical papers (see,
e.g., Refs.~[\onlinecite{Caroli,Maki}] where the situation near
the upper critical field $H_{c2}$ was discussed). In dirty
superconductors, the electron contribution to the thermal
conductance along the vortices is proportional to the area
occupied by the cores $\kappa(B)\simeq(B/H_{c2})\kappa_N$, where
$\kappa_N$ is the electron thermal conductance in the normal
state.~\cite{Bardeen,Bardeen-Rickayzen-1959,Kopnin/heat} In clean
superconductors this simple estimate fails to describe the
experimental data~\cite{Lowell,vinen}: the thermal conductance
along the well-separated vortices in $s$-wave superconductors
appears to be two orders of magnitude smaller than predicted. This
discrepancy was first explained in Ref.~[\onlinecite{vinen}] as
resulting from the very small group velocity of the Caroli--de
Gennes--Matricon (CdGM) states~\cite{degen} localized inside the
cores. Ideas of Ref.~[\onlinecite{vinen}] were cast into a
quantitative theoretical framework in Ref.~[\onlinecite{KMV03}]
where the vortex line was treated as a long ballistic quantum
channel. It was shown that the electronic thermal transport
through a vortex can be described within the framework of the
Landauer approach in a manner similar to that used for heat
transport in normal conductors\cite{Butcher90,Houten92}. The
vortex heat conductance can be expressed in terms of the ``heat
conductance quanta'' as a certain amount of heat carried by each
quantized conducting mode\cite{Rego99}. This yields $\kappa _{\rm
L} =\kappa_0 N_{\rm L}$, where the quantum $\kappa_0=\pi
T/(3\hbar)$ is the universal heat conductance per conducting mode
in a normal metal (we use the units with $k_B=1$), and $N_{\rm L}$
is the effective number of conducting modes in the vortex (for
both spin projections of electron). Quantization of the heat
conductance in normal conductors has been observed in several
experiments (see, e.g., Refs.~[\onlinecite{Houten92,Chiatti06}]).

To estimate the number of modes $N_{\rm L}$ we summarize the main
properties of the vortex core spectrum. The CdGM spectrum,
$\epsilon _\mu$, is a function of the quantized (half-integer)
angular momentum $\mu=bk_\perp$, where $b$ is the impact
parameter, $k_\perp=\sqrt{k_F^2-k_z^2}$, $k_F$ is the Fermi
momentum, and $k_z$ is the momentum projection on the vortex axis.
As $\mu$ changes from $-\infty$ to $+\infty$, $\epsilon _\mu$
varies from $-\Delta_0$ to $+\Delta_0$ crossing zero when $\mu$
changes its sign.  Here $\Delta_0$ is the gap value far from the
vortex axis. At small energies, $|\varepsilon|\ll\Delta_0$, the
spectrum is $\epsilon _\mu \simeq -\mu \Delta_0/(k_\perp\xi)$,
where $\xi=\hbar V_F/\Delta_0$ is the superconducting coherence
length, and $V_F =\hbar k_F/m$ is the Fermi velocity. The
interlevel spacing $\omega_0\simeq\Delta_0/(k_F\xi)$ taken at
$k_z=0$ determines the excitation minigap $\omega_0/2$. For
temperatures $\omega_0\ll T\ll \Delta_0$, the number of modes is
thus~\cite{KMV03} $N_{\rm L}\sim T/\omega_0$, and the heat
conductance of one vortex becomes
\begin{equation}
 \label{Landauer-cond}
 \kappa_{\rm L} \sim T^2/\hbar\omega_0 \ .
\end{equation}
This estimate can also be obtained using the Sharvin conductance
with the group velocity $v_g=\partial\epsilon _\mu /\hbar\partial
k_z\sim\epsilon _\mu /\hbar k_F$ rather than $v_z\sim v_F$ as in a
normal tube. Comparing $N_{\bf L}$ with the corresponding
Sharvin's number of channels in the normal wire of radius $\xi$,
$N_{\rm Sh}\sim (k_F\xi)^2$, one sees that $N_{\rm L}/N_{\rm
Sh}\sim v_g/v_F\sim (T/T_c)(k_F\xi )^{-1}\ll 1$.  At extremely low
temperatures the minigap in the spectrum suppresses the heat
transport exponentially: $\kappa\propto\exp(-\omega_0/2T)$ when
$T\ll\omega_0$.

The qualitative estimate of Eq.\ (\ref{Landauer-cond}) applies to
the heat conductance along the vortex core in the samples infinite
in the lateral direction, but as we will see below does not
generally work for samples finite in the lateral direction. Recent
remarkable progress in experimental technique made it possible
direct heat transport measurements in mesoscopic
normal-superconducting heterostructures\cite{Therm-review} of
sizes comparable to the superconducting coherence length, opening,
in particular a new route for the analysis of quantum heat
transport through individual vortices in mesoscopic
superconductors whose exotic vortex states are in the focus of
current research~\cite{Geim}. This calls for a comprehensive
quantitative theory of the heat transport in mesoscopic
superconductors.

In this paper we develop a theory for an electronic heat
conductance along the normal channels (vortex cores) in the
superconductors of the arbitrary lateral size and shape, focusing
on the specific features that appear in the mesoscopic cylinders
with diameters comparable to the coherence length $\xi$. We show
that in {\it finite} superconductors, especially those with the a
transverse size of the order of a few coherence lengths $\xi$, the
estimate Eq.\ (\ref{Landauer-cond}) is no longer valid. The
quantum mechanics of quasiparticle excitations in mesoscopic
superconductors turns out to be very sensitive to the boundary
effects.  In particular, the interplay between the Andreev
reflection at the core boundary and the normal scattering of
quasiparticles at the sample surface results in an oscillatory
behavior of the energy levels as functions of the Fermi momentum
and the sample lateral size~\cite{PRL-2005}. These conclusions
have been recently confirmed also for a graphene channel in
contact with a superconducting environment\cite{Titov06}. For a
vortex in a mesoscopic sample, the amplitudes of level
oscillations can well exceed the CdGM interlevel spacing. In this
case, the new quasiparticle modes propagating along the vortex
core appear which leads to a dramatic increase in the effective
number of the conducting channels, $N_{\rm L}$, as compared to
that in Eq.\ (\ref{Landauer-cond}). We present a theoretical
analysis of the single-particle transport mechanisms in the mixed
state of mesoscopic superconductors taking into account normal
reflections of the quasiparticle waves at the superconductor
boundaries. We concentrate on the low-temperature $T\ll\Delta_0$
regime for the ballistic samples with the mean free paths, both
elastic and inelastic, being much larger than the sample
dimensions. We further investigate the effects of the surface
roughness on the excitation spectrum and, accordingly, on the heat
transport.  We show that the significant increase in the number of
conducting modes caused by the spectrum oscillations is a generic
feature of a broad class of surface imperfections.

The paper is organized as follows. In Section \ref{Sec:spectrum}
we introduce the model and consider the basic equations describing
the quasiparticle quantum mechanics for a vortex line placed in a
mesoscopic superconducting cylinder with the surface of an
arbitrary shape. A Landauer-type analysis of the heat transport
due to the quasiparticle modes propagating along the vortex core
is carried out in the Section \ref{Sec-Landauer}. In this section
we also present the results of numerical calculations of the
vortex heat conductance in an ideal mesocopic cylinder. Then in
the Section \ref{Sec-Roughness} we investigate the effect of the
surface roughness on the quasiparticle spectrum.

\section{Quasiparticle spectrum for a vortex line in a
mesoscopic superconductor}
\label{Sec:spectrum}

\subsection{Semiclassical equations}

In this section we consider a vortex placed in a superconducting
mesoscopic cylinder with an arbitrary $(x,y)$ cross section and
derive an eikonal approximation of the Bogoliubov--de Gennes (BdG)
equation that allows us to obtain the vortex-core states in such
samples. Let the vortex line be parallel to the cylinder axis $z$.
We choose the origin of the polar coordinate system $(r,\, \theta
,\, z)$ of the $(xy)$ plane at the vortex center. The surface of
the sample is then defined by the equation $r=R(\theta)$. Since
the momentum component along the vortex axis is a conserved
quantity, we choose the particle-- and hole--like parts of the
wave function as $(U,V)=(u,v)\exp(ik_zz)$ respectively. The BdG
equations have the form
\begin{equation}
\left(\begin{array}{cc} \hat H_0(\hat {\bf p})
& \Delta(x,y) \\
\Delta^*(x,y) & -\hat H_0^*(\hat {\bf p})
\end{array}\right)
\left(\begin{array}{c} u \\ v \end{array} \right) =\varepsilon
\left(\begin{array}{c} u \\ v\end{array} \right)\,. \label{BdG}
\end{equation}
Here
\[
\hat H_0(\hat {\bf p}) =\displaystyle\frac{\mathstrut
1}{\mathstrut 2m} \left(\hat{\bf p}-\frac{e}{c}{\bf
A}\right)^2-E_{\perp}\ ,
\]
$\hat {\bf p}=-i\hbar\nabla$, $E_{\perp}=E_F-\hbar^2
k_z^2/(2m)=\hbar^2 k_\perp^2/(2m)$, and $E_F$ is the Fermi energy.
The order parameter profile $\Delta (x,y)$ is homogeneous in the
$z$ direction,
\[
\Delta (x,y) =\Delta_0\delta_v (r) e^{i\varphi+i\tilde \varphi} \
.
\]
Here $\delta _v(r)$ is the normalized order parameter magnitude
for a vortex centered at $r=0$, such that $\delta _v(r)=1$ for
$r\rightarrow \infty$, $e^{i\varphi }=(x+iy)/r$ is the vortex
phase factor, and $\tilde \varphi$ is the part of the order
parameter phase induced by the boundary effects; $\tilde\varphi$
having no singularities inside the superconductor. For simplicity
we start our analysis with the case of the weak external magnetic
field and the extreme type-II superconductors where the vector
potential ${\bf A}$ can be neglected. The effects of the magnetic
field are discussed later in Section \ref{mf}.

Let us introduce a momentum representation:
\begin{equation}
\left(u \atop v \right) = \frac{1}{(2\pi\hbar)^2}\int d^2p\,
e^{i{\bf p}{\bf r}/\hbar} \hat \Psi({\bf p}) \ , \label{psi_p}
\end{equation}
where ${\bf p} = p(\cos\theta_p, \sin\theta_p)= p {\bf p}_0$. The
unit vector ${\bf p}_0$ parametrized by the angle $\theta _{p}$
defines the trajectory direction in the $(x,y)$ plane. Within the
quasiclassical approach the wave function in the momentum
representation assumes the form:
\begin{equation}
\hat\Psi({\bf p})=
\frac{1}{k_\perp}\int\limits_{-\infty}^{+\infty} ds\, e^{-i(|{\bf
p}|-\hbar k_\perp)s/\hbar} \hat\psi(s,\theta_p) \ .\label{st}
\end{equation}
We look for solutions with $|{\bf p}|\approx\hbar k_\perp$. For
the slowly varying part of the wavefunction $\hat\psi$ we obtain
the quasiclassical (Andreev) equation along a trajectory specified
by an orientation angle, $\theta_p$, and an impact parameter, $b$,
with respect to the vortex center. The distance along the
trajectory is $s=r \cos(\theta_p-\theta)$. The wavefunction for
${\bf r}=r(\cos\theta,\sin\theta)$ is derived from Eqs.
(\ref{psi_p}), (\ref{st}):
\begin{equation}
\left(u \atop v \right) =
\frac{1}{2\pi}\int\limits_{0}^{2\pi}d\theta_p\, e^{i k_\perp r
\cos(\theta_p-\theta)} \hat\psi(r \cos(\theta_p-\theta),\theta_p)
\ .
\end{equation}
The wave functions should vanish at the sample surface:
\begin{equation}
\frac{ 1}{2\pi} \int\limits_{0}^{2\pi}d\theta_p e^{i k_\perp
R(\theta) \cos(\theta_p-\theta)} \hat\psi(
R(\theta)\cos(\theta_p-\theta),\theta_p) =0 \ . \label{bcond1}
\end{equation}

We introduce the angular-momentum expansion
\[
\hat \psi = \sum_\mu e^{i\mu\theta_p+i\hat\sigma_z
[\theta_p+\tilde \varphi (s,\theta_p)]/2}\hat g_\mu (s,\theta_p) \
,
\]
where $ \mu= -k_\perp b =n+ 1/2 $, and $n$ is an integer. The
function $\hat g_\mu$ satisfies the equation: $\hat H \hat g_\mu =
\varepsilon \hat g_\mu$, where
\begin{eqnarray}
\hat H &=& -i\hbar V_\perp \hat\sigma_z\frac{\partial}{\partial
s}+ \Delta_0 \delta_v(r)
\left[ \hat\sigma_x s- \hat\sigma_y b \right]/r \nonumber \\
&&+\hbar k_\perp \tilde v_s (s,b, \theta_p)\ ,
\end{eqnarray}
$V_\perp =\hbar k_\perp /m$, $\hat\sigma_x, \hat\sigma_y,
\hat\sigma_z$ are the Pauli matrices, $r=\sqrt{s^2+b^2}$, and
\[
\tilde v_s=\frac{\hbar}{2m}\frac{\partial\tilde \varphi }{\partial
s} =\frac{\hbar}{2m}{\bf p}_0\cdot \nabla\tilde \varphi
\]
is the projection on the momentum direction of the superfluid
velocity component induced by the boundary effects; we neglect
this velocity assuming for simplicity that it is small as compared
to the critical velocity at distances shorter than $\xi$ from the
vortex axis. The function $\hat g_\mu$ is thus independent of
$\theta_p$: $\hat g_\mu =\hat g_\mu(s)$. This assumption does not
change the main results of our work.

The main contributions to the integral in Eq. (\ref{bcond1}) come
from the vicinities of the stationary phase points, $
\sin(\theta_p-\theta)=\mu/k_\perp R(\theta) $. We further assume
that the r.h.s. of this equation is small, i.e. that the impact
parameters for all trajectories contributing to the solution are
much less than the system size. As a result we obtain two stationary
points: $ \theta_p \simeq \theta + \mu/k_\perp R(\theta)$ and $
\theta_p \simeq \pi +\theta - \mu/k_\perp R(\theta)$. Furthermore,
we neglect the corrections $\mu/k_\perp R$ in the above expressions
for stationary angles; this is valid provided  (i) the additional
phase shifts along trajectories caused by the finite impact
parameters are small $k_\perp b^2/R \ll 1$, and (ii) $\hat g_\mu
(s)$ is a slow function on the atomic length scale.

Near the stationary angle we can write
$\cos(\theta_p-\theta)\simeq 1-(\theta_p-\theta)^2/2$ and
$\cos(\theta_p-\theta)\simeq -1+(\theta_p-\theta-\pi)^2/2$,
respectively. The integral in Eq. (\ref{bcond1}) becomes
\begin{eqnarray*}
&&\int\limits_{0}^{2\pi}d\theta_p
 e^{i k_\perp R(\theta) \cos(\theta_p-\theta)}
\hat\psi[ R(\theta)\cos(\theta_p-\theta),\theta_p]
\\
&=&\sqrt{\frac{2\pi}{k_\perp R(\theta)}}\left[ e^{i k_\perp
R(\theta)-i\pi/4} \hat\psi( R(\theta),\theta)\right. \\
&& + \left. e^{-i k_\perp R(\theta)+i\pi/4} \hat\psi(
-R(\theta),\theta+\pi)\right] \ .
\end{eqnarray*}
Thus, the boundary condition Eq. (\ref{bcond1}) takes the form:
\begin{equation}
\hat\psi( R(\theta),\theta) = - e^{-2i k_\perp R(\theta)+i\pi/2}
\hat\psi( -R(\theta),\theta+\pi) \ . \label{bcond2}
\end{equation}

To find the function $\hat g_\mu$ we can use the results of
Ref.~\cite{KMV03}. Introducing
\begin{equation}
\hat g_\mu= e^{\hat\sigma_z\left(i\frac{\pi}{2}-i\frac{\pi}{4}
{\rm sign} b - \frac{i}{2}\arctan\frac{s}{b} \right)}\hat w_\mu
\end{equation}
we obtain:
\begin{equation}
-i\hbar V_\perp \hat\sigma_z\frac{\partial}{\partial s}\hat w_\mu+
\hat\sigma_x \Delta_0 \delta_v \hat w_\mu = \left(\varepsilon +
\frac{\hbar V_\perp}{2}\frac{b}{s^{2}+b^{2}}\right)\hat w_\mu
 \ .
\end{equation}

The two linearly independent solutions for low energies are
\cite{KMV03}:
\begin{eqnarray*}
\hat w_{1\mu} (s) &=&
e^{-D(s)/2}e^{\frac{i}{2}\arctan(s/b)\hat\sigma_z+i\hat\sigma_z[{\rm
sign}(b)-1]\pi/4}\\
&&\times
\sqrt{\frac{e^{D-|D|}+\gamma^2}{(1+\gamma^2)(e^{2D}+\gamma^2)}}
\left[e^D+i\gamma\hat\sigma_z\right]\hat \lambda \ ,
\end{eqnarray*}
and $ \hat w_{2\mu} (s) ={\rm sign} (b)\, \hat w_{1\mu}^* (-s) $.
We denote
\[
\hat \lambda = \left(1\atop 1\right) \ ,
\]
\begin{equation}
D(s)=\frac{2\Delta_0}{\hbar V_\perp}\int\limits_0^s
\delta_v(s^\prime) ds^\prime \ , \;  \Lambda =
\frac{2\Delta_0}{\hbar V_\perp} \int\limits_0^\infty e^{-D(s)} ds
\ , \label{DL-defin}
\end{equation}
\begin{equation}
\gamma (\mu) =\frac{2}{\hbar V_\perp} \left[\epsilon (\mu) -
\varepsilon\right]\int\limits_0^\infty e^{-D(s)} ds \ .
\end{equation}
Here
\begin{equation}
\epsilon (\mu)= -\frac{2\Delta _0^2\mu}{\hbar  k_\perp
V_\perp \Lambda} \int_0^{\infty}\frac{\delta (s)}{s} e^{-D(s)}\,
ds \simeq -\frac{\mu\omega _0 k_F}{k_\perp}
\end{equation}
is the CdGM energy spectrum for an infinite sample.

We take the function $\hat g_\mu$ as a sum
\[
\hat g_\mu = c_{1\mu} \hat G_{1\mu} +c_{2\mu} \hat G_{2\mu}
\]
of two linearly independent solutions
\begin{eqnarray*}
\hat G_{1\mu} &=& e^{+\hat\sigma_z\left(i\frac{\pi}{2}-
i\frac{\pi}{4} {\rm sign} b - \frac{i}{2}\arctan\frac{s}{b}
\right)} \left(\hat w_{1\mu}+\hat w_{2\mu}\right)/2 \ ,
\\
\hat G_{2\mu} &=&
e^{+\hat\sigma_z\left(i\frac{\pi}{2}-i\frac{\pi}{4} {\rm sign} b -
\frac{i}{2}\arctan\frac{s}{b} \right)}\left(\hat w_{1\mu}- \hat
w_{2\mu}\right)/i\gamma \ .
\end{eqnarray*}
The characteristic length scale of the function $\hat g_\mu$ is of
the order of $\xi$. Assuming $\gamma\ll 1$ these two solutions can
be rewritten in the form:
\begin{eqnarray*}
\hat G_{1\mu} &=& e^{i\hat\sigma_z \pi/4} \left(e^{-|D(s)|/2} -i
\, {\rm sign}s\, \frac{\gamma}{2}\hat\sigma_z e^{|D(s)|/2}\right)
\hat \lambda \ ,
\\
\hat G_{2\mu} &=&e^{i\hat\sigma_z \pi/4}e^{|D(s)|/2} \hat\sigma_z
\hat \lambda  \ .
\end{eqnarray*}
Substituting all the above expressions into the boundary condition
and taking account of the fact that for small $\mu$ we have
$\tilde\varphi (R,\theta)\simeq \tilde\varphi (-R,\theta+\pi)$, we obtain
\begin{eqnarray*}
\sum_\mu e^{i\mu\theta}\left\{ c_{1\mu}\! \left(e^{-D_\theta/2} -
i\frac{\gamma}{2}\hat\sigma_z e^{D_\theta/2}\right)\! + c_{2\mu}
e^{D_\theta/2}\hat\sigma_z -\right.&&   \\
\left. -e^{i\pi\mu -i\alpha _\theta}\! \left[ c_{1\mu}\!
\left(\hat\sigma_z e^{-D_\theta/2} + i\frac{\gamma}{2}
e^{D_\theta/2}\right)\! + c_{2\mu} e^{D_\theta/2} \right]\right\}
\hat \lambda \! &=&\! 0
\end{eqnarray*}
where $\alpha _\theta =2k_\perp R(\theta)$, $D_\theta
=D(R(\theta))$. These expressions hold as long as
$e^{-D_\theta}\ll 1$. Let us now introduce the angular functions
\[
c_j(\theta) = \sum_\mu e^{i\mu\theta} c_{j\mu} \ , \; c_{j\mu}
=\frac{1}{2\pi} \int\limits_0^{2\pi} e^{-i\mu\theta} c_j(\theta)
d\theta \ .
\]
They satisfy
\begin{eqnarray*}
\label{e11} -\frac{i}{2}\hat\gamma c_1(\theta)+c_2 (\theta)&=&
e^{-D_\theta-i\alpha_\theta}c_1(\theta+\pi) \ , \\
\label{e12} \frac{i}{2}\hat\gamma c_1(\theta+\pi)+c_2
(\theta+\pi)&=& e^{-D_\theta+i\alpha_\theta}c_1(\theta) \ .
\end{eqnarray*}
$\gamma$ is now considered as a function of $\mu$ which is
transformed into an operator
\[
\hat\gamma \equiv \hat \gamma (\hat \mu ) =
\gamma\left(-i\frac{\partial}{\partial\theta}\right) \ .
\]
Since $c_j(\theta+2\pi)=-c_j(\theta)$ while $\alpha _\theta$ and
$D_\theta$ are $2\pi$-periodic we obtain
\begin{eqnarray*}
c_2(\theta+\pi)&=&\frac{1}{2}\left(e^{-D_\theta+i\alpha_\theta}-
e^{-D_{\theta+\pi}-i\alpha_{\theta+\pi}}\right) c_1(\theta)\ ,
\\
i\hat\gamma c_1(\theta+\pi)&=&\left(e^{-D_\theta+i\alpha_\theta}+
e^{-D_{\theta+\pi}-i\alpha_{\theta+\pi}}\right)c_1(\theta) \ .
\end{eqnarray*}
This set of equations demonstrates that the normal reflection at
the surface couples the two trajectories going in the opposite
directions that are described by two independent functions
$c_1(\theta)$ and $c_1(\theta+\pi)$. Our derivation uses
trajectories with small impact parameters and, thus, assumes that
these equations are valid provided the angular harmonics
$c_{1\mu}$ vanish for $|\mu|> {\rm min} \{k_\perp\xi,
\sqrt{k_\perp R} \}$.

Following Ref.~[\onlinecite{larkin}] it is convenient to introduce
a two-component function
\begin{equation}
\hat c (\theta) =\left(c_1(\theta) \atop c_1(\theta+\pi)\right)\ ,
\end{equation}
defined for $0<\theta<\pi$ with the boundary condition $ \hat c
(\theta+\pi) =i\hat\sigma_y \hat c (\theta)$. The final equation
for $\hat c$ reads
\begin{equation}
\hat h \hat c =\varepsilon\hat c \ . \label{eikonalequation1}
\end{equation}
where the effective Hamiltonian is $\hat h=\hat h_0+\hat V$,
\begin{equation}
\hat h_0 = \epsilon
\left(-i\frac{\partial}{\partial\theta}\right)\ , \; \hat V =
-\Lambda^{-1}\Delta_0 \left( \begin{array}{lr} 0& v_\theta
\\
v_\theta^* & 0 \end{array} \right) \label{eikonalequation2}
\end{equation}
and
\begin{equation}
v_\theta =i\left(e^{-D_\theta-i\alpha_\theta}+
e^{-D_{\theta+\pi}+i\alpha_{\theta+\pi}}\right)\ .
\label{eikonalequation3}
\end{equation}
Equations (\ref{eikonalequation1})--(\ref{eikonalequation3}) are
the central result of this Section. They provide a semiclassical
description of vortex core excitations taking into account the
normal scattering processes at the sample surface of a general
shape. The normal scattering is introduced by the potential $\hat
V$ through the factors $e^{i\alpha _\theta}$ which result in
mesoscopic oscillations of the spectrum as functions of momentum
and sample size \cite{PRL-2005}. We employ these equations to
study the effects of surface imperfections further in Section
\ref{Sec-Roughness}.

\subsection{Spectrum of an ideal cylinder}

\begin{figure}[hbt]
\centerline{\includegraphics[width=1.0\linewidth]{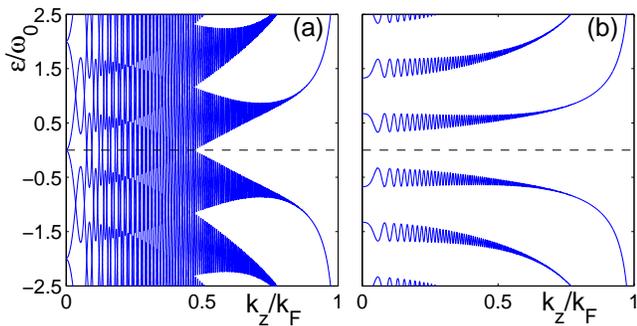}}
\caption{\label{fig:spectra} The spectra of quasiparticles for a
vortex in the cylinder with $R/\xi=3.5$ (a) and $R/\xi=4.5$ (b).
The gap is approximated by Eq.~(\ref{Core-model}) with
$\xi_v=\xi$, $k_F\xi=200$.}
\end{figure}

For an ideal cylinder with a constant radius $R(\theta)=R$, the
solution of equations
(\ref{eikonalequation1})--(\ref{eikonalequation3}) were obtained
in Ref.~[\onlinecite{PRL-2005}]; we reproduce them here for the
sake of completeness and convenience. The wavefunction is
\begin{equation}
\hat c_\mu =\frac{e^{i\mu \theta}}{\sqrt{2\pi}} \left( 1\atop
e^{i\pi\mu}\right) \ , \label{egenfunct-ideal}
\end{equation}
where $\mu =n+1/2$, which gives the energy spectrum
\cite{PRL-2005}
\begin{equation}
\varepsilon_\mu =\epsilon (\mu)+\Delta_0\frac{2e^{-D(R
)}}{\Lambda} \cos(2k_\perp R -\pi\mu+\pi/2)\
,\label{spectrum-cylinder}
\end{equation}
Using a simple vortex core model \cite{PRL-2005}
\begin{equation}
\delta_v(r)=r/\sqrt{r^2+\xi_v^2}\, ,\label{Core-model}
\end{equation}
where $\xi_v\sim\xi$ is the size of the vortex core, one obtains
\begin{eqnarray}
\frac{\varepsilon_\mu }{\Delta_0}&=&-\frac{\mu}{k_\perp\xi_v}
\frac{\mathcal{K}_0(a)}{\mathcal{K}_1(a)} \nonumber \\
&&+\frac{2e^{-a\sqrt{R^2 /\xi_v^2+1}}}
{a\,\mathcal{K}_1(a)}\cos\left[2k_\perp R
-\pi\mu+\frac{\pi}{2}\right]\ .\label{spectrum-model}
\end{eqnarray}
Here $a=2(\xi_v/\xi)k_F/k_\perp$ and $\mathcal{K}_n$ is the
McDonald function. Typical energy spectra of
Eq.~(\ref{spectrum-model}) for a vortex in a mesoscopic cylinder
are shown in Fig.~\ref{fig:spectra}.
The minigap in the spectrum exists only for $R>R_c$, where the
critical radius is
\begin{equation}
R_c=\frac{\xi}{2}\sqrt{\ln^2\left[\frac{2k_F\xi}{
\mathcal{K}_0(2\xi_v/\xi)}\right]-\frac{4\xi_v^2}{\xi^2}} \simeq
\frac{\xi}{2}\ln\left[\frac{2k_F\xi}{
\mathcal{K}_0(2\xi_v/\xi)}\right] \ .\label{Rc}
\end{equation}
The interlevel spacing $\omega_0$ for CdGM states is
\begin{equation}
\omega_0=\frac{\Delta_0}{k_F\xi_v}\frac{\mathcal{K}_0(2\xi_v/\xi)}
{\mathcal{K}_1(2\xi_v/\xi)}\,.\label{omega0}
\end{equation}
Taking for simplicity $\xi_v=\xi$  we get $\omega_0\simeq 0.8143
\Delta_0/k_F\xi$.

\subsection{Magnetic-field tuning of the spectrum oscillations \label{mf}}

In this section we consider the influence of an external magnetic
field on the spectral characteristics and show that the
quasiparticle spectrum can be tuned by changing the magnetic field
${\bf H}=H{\bf z}_0$ applied to the sample.

With the account of the magnetic field, the CdGM energy is
renormalized\cite{BrunHansen}, $\varepsilon _\mu \rightarrow
\varepsilon_\mu +\mu \omega_c /2$ where $\omega_c=eH/mc$ is the
cyclotron frequency. However, the most important is the effect of
the magnetic field on the gap profile. Far from the core it
becomes
\begin{equation}
\delta_v\simeq 1 - \frac{\pi^2\xi^2 H^2 r^2}{2\phi_0^2} \ .
\label{delta-H}
\end{equation}
where $\phi _0= \pi \hbar c/e$ is the magnetic flux quantum. We
assume that the supercurrent density at the sample edge is less
than the critical current so that the role of states bound to the
edge \cite{nature} can be neglected for low energies. The change
in the gap profile results in an additional small renormalization
of the minigap and, what is more important, it also affects the
amplitude of level oscillations determined by the factor
$\exp(-D(R))$. Indeed,we find from Eqs. (\ref{DL-defin}),
(\ref{delta-H})
\[
D(R)=\frac{k_F}{k_\perp}\left(\frac{2 R }{\xi } -\frac{\xi h^2
}{3R}\right)  \ ,
\]
where $h=H/H_0$ and $ H_0=\phi_0/\pi R^2 $. The field $H_0$ is of
the order of the critical field $ H_1=(\phi_0/\pi R^2) \ln (R/\xi)
$ at which the vortex becomes energetically favorable in the
sample. The ratio of the oscillation amplitudes taken, e.g., for
$h=3$ and $h=1$ is given by the expression:
\[
\frac{e^{-D(R,h=3)}}{e^{-D(R,h=1)}} =
\exp\left(\frac{8k_F\xi}{3k_\perp R}\right)
 \ .
\]
For $k_\perp=k_F$ and $R/\xi=4$ this estimate gives us a factor
$\sim 2$, which shows that the amplitude of oscillations can be
varied significantly by changing the magnetic field in a
reasonable range. The critical radius $R_c$ at which zero energy
modes appear is determined by the condition $D(R_c,H)\simeq
\ln(k_F\xi)$ and does also depend on $H$. To conclude here, one
observes that the energy level oscillations and the number of
propagating quasiparticle modes depend strongly on the external
magnetic field and thus can be effectively controlled in the
experiments.

\section{The Landauer approach to the vortex heat conductance}
\label{Sec-Landauer}

\subsection{Semiclassical scheme}\label{subsec-semiclassics}

Now we generalize the Landauer approach of
Ref.~[\onlinecite{KMV03}] developed for the electronic heat
transport in superconductors with infinite transverse dimensions
to mesoscopic samples and derive the along-the-vortex heat
current. A general expression for the energy current along $z$ is
\begin{eqnarray}
I_{{\cal E}} \!&=&\!\!\! \int  d^2r \sum _\mu \int \frac{d
k_z}{\pi m}\, \left[\varepsilon _\mu \, u^*_{\mu k_z} \left(\hbar
k_z-\frac{e}{c}A_z\right)u_{\mu k_z} n(\varepsilon_\mu )\right.
\nonumber \\
&&\!\! -\left. \varepsilon_\mu \, v^*_{\mu k_z} \left(\hbar k
_z+\frac{e}{c}A_z\right)v_{\mu k_z} \left[1-n(-\varepsilon_\mu
)\right] \right] \ . \label{encurrent/def}
\end{eqnarray}
Particles with the distribution $n(\varepsilon)$ carry the energy
$+ \varepsilon$ while the holes with the distribution
$1-n(-\varepsilon)$ carry the energy $-\varepsilon$. If the
electrodes are in equilibrium, $1-n(-\varepsilon) =n(\varepsilon)$
at each electrode. The sum is taken over the states with
$\varepsilon _\mu >0$. Using the BdG equations one can derive the
important identity for the wave functions of the bound states
$u_{\mu k_z}, \, v_{\mu k_z}$ belonging to the eigenvalues
$\varepsilon _\mu$ of a given momentum $k_z$ along the
$z$--axis\cite{KMV03}
\begin{eqnarray*}
\int \left[ u^*_{\mu k_z}(
 k_z-\frac{e}{\hbar c}A_z)u_{\mu
 k_z} -  v^*_{\mu k_z}(
 k_z+\frac{e}{\hbar c}A_z )v_{\mu k_z}\right] d^2r \nonumber \\
 =\frac{m}{\hbar^2}\frac{\partial \varepsilon _{ \mu }}{\partial k_z}
\ .
\end{eqnarray*}
For the CdGM case this identity reveals a huge cancellation in its
left hand side where each term is by the factor $k_F\xi$ larger
than the right hand side. Within the quasiclassical approximation
the l.h.s. vanishes for an infinite vortex line as a direct
consequence of an approximate electron-hole symmetry. Note that
such a cancellation does not occur in a finite-thickness
slab\cite{KMV03} where the CdGM states are not truly localized.
Making use of this identity reduces the expression for the heat
current to
\begin{equation}
I_{{\cal E}}=\sum_\mu\int\limits_{-k_F}^{+k_F}
 \varepsilon_\mu n(\varepsilon_\mu )\,
 \frac{\partial\varepsilon_\mu}{\partial k_z}\,
 \frac{dk_z}{\pi\hbar} \ . \label{heatcurrent}
\end{equation}
Equation~(\ref{heatcurrent}) shows that, in a contrast to the
electrical current, the energy flow is determined by the group
velocity of excitations and this is why the conductance in
Eq.~(\ref{Landauer-cond}) is much smaller than $\kappa_{\rm
Sh}=\kappa_0N_{\rm Sh}$, where $\kappa_0=\pi T/(3\hbar)$ is the
heat conductance quantum defined in the Introduction.

The quasiparticles in the injector lead (left lead) having
energies coinciding with the core-state levels penetrate into the
core and are then transmitted through the vortex to the right lead
with a probability $P_{\mu, k_z}$. In the same manner as it was
done for the charge transport in mesoscopic
conductors\cite{Butcher90,Houten92,Landauer} we conclude that
excitations with a positive group velocity
$v_g=\partial\varepsilon_\mu/\hbar\partial k_z$ have the
distribution $n_L P_{\mu, k_z} $ where $n_L$ is the distribution
they had in the left lead. Similarly, excitations with a negative
group velocity possess the distribution $n_R P_{\mu, k_z} $ where
$n_R$ is their distribution in the right lead. Since
$\varepsilon_\mu$ is an even function of $k_z$ we obtain
\begin{equation}
 I_{{\cal E}}=\sum _\mu \int\limits_{0}^{k_F}  \varepsilon _\mu \left[
 n_L(\varepsilon_\mu )-n_R(\varepsilon_\mu )\right]P_{\mu, k_z}\, \left|
 \frac{\partial \varepsilon _\mu }{\partial k_z}\right|\,
 \frac{d k_z}{\pi\hbar} \ .
 \label{heatcurrent-Land}
\end{equation}
Defining the heat conductance as
\[
I_{{\cal E}}=\kappa (T_L-T_R)
\]
where $T_{L,R}$ are the temperatures of the left and right
electrodes respectively, one finds
\begin{equation}
\label{heat-cond-Land-estim}
 \kappa =-\frac{1}{\pi T\hbar }\sum_\mu\int\limits_0^{k_F}
 \varepsilon_\mu ^2 \frac{d n(\varepsilon_\mu)}{d\varepsilon _\mu}
P_{\mu, k_z} \left|\frac{\partial\varepsilon_\mu}{\partial
k_z}\right|\,dk_z\,.
\end{equation}
We assume that $T_L-T_R\ll T_{L,R}\approx T $. Using the
conductance quantum $\kappa_0$, one can introduce the effective
number of transverse modes propagating along the vortex line
$N_{eff}=\kappa/\kappa_0$. Later in this section we neglect the
scattering of excitations at the interface between the leads and
the superconductor assuming that they are transmitted freely
through the vortex, i.e. with $P_{\mu, k_z}=1$, deferring the
effects of scattering till Section \ref{subsec-Exact}.

For an infinite sample $R\gg R_c$ in the limit $\omega_0\ll T\ll
\Delta_0$ the sum in Eq.~(\ref{heat-cond-Land-estim}) can be
replaced with the integral over $d\mu$ for the states with
positive energies.  Since $dn/d\varepsilon
=-(4T)^{-1}\cosh^{-2}(\varepsilon/2T)$ we find
\[
\kappa =\frac{9\zeta(3)}{2\pi}\frac{T^2}{\hbar \omega_0} \ .
\]
Here $\zeta(n)$ is the Riemann zeta-function. Therefore
\[
 N_{eff}^{(\infty)}(T)=\frac{27\zeta(3)}{2\pi ^2}\frac{T}{\omega_0}
\sim k_F\xi\frac{T}{T_c} \ ,
\]
which coincides with the estimate discussed in the introduction
(see Eq.~(\ref{Landauer-cond})). At extremely low temperatures,
$T\ll\omega_0$:
\begin{equation}
\label{Neff-T=0-R>Rc}
 N_{eff}^{(\infty)}(T)=\frac{3}{4\pi^2}\left(\frac{\omega_0}{T}\right)^2
 \exp\left(-\frac{\omega_0}{2T}\right)\ .
\end{equation}

\begin{figure}[hbt]
\centerline{\includegraphics[width=1.0\linewidth]{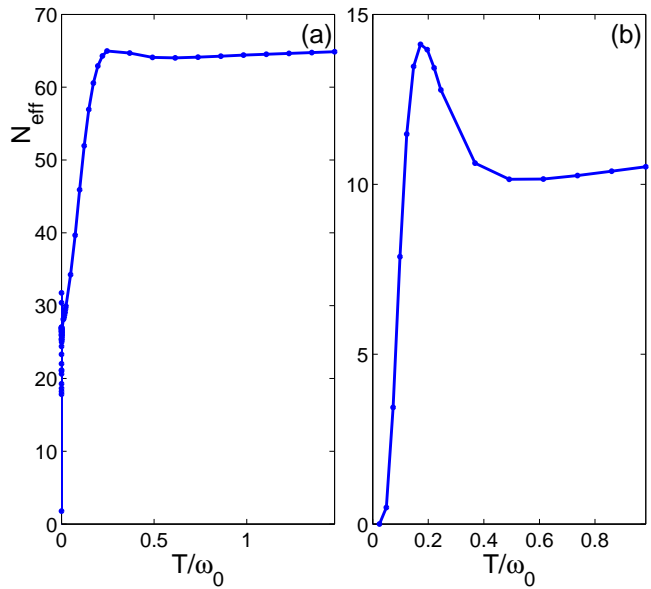}}
\caption{\label{fig:Neff-T-analyt} The temperature dependence of
the effective number of modes $N_{eff}(T)$ calculated using
Eq.~(\protect\ref{heat-cond-Land-estim}) with $P_{\mu, k_z}=1$ for
two cylinders with $R/\xi=3.5$ (a) and $R/\xi=4.5$ (b). The gap is
approximated by Eq.~(\protect\ref{Core-model}) with $\xi_v=\xi$,
$k_F\xi=200$.}
\end{figure}
\begin{figure}[hbt]
\centerline{\includegraphics[width=1.0\linewidth]{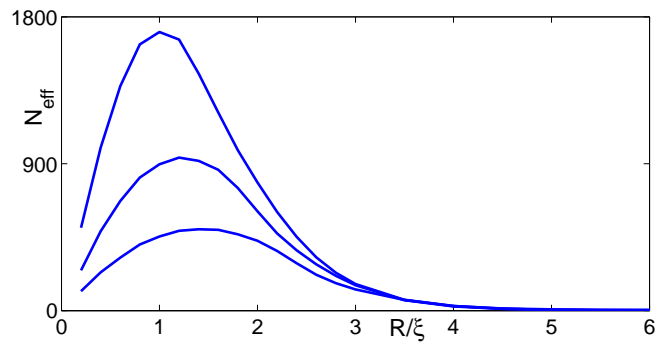}}
\caption{\label{fig:Neff-R} The effective number of modes
$N_{eff}$ according to Eq.~(\protect\ref{heat-cond-Land-estim})
with $P_{\mu, k_z}=1$ as a function of the cylinder radius for
various temperatures $T/\omega_0=0.3, 1.0, 3.0$ (from bottom to
top). The gap is approximated by Eq.~(\protect\ref{Core-model})
with $\xi_v=\xi$, $k_F\xi=200$.}
\end{figure}

In a mesoscopic cylinder with $R\lesssim R_c$ there appear new
conducting modes due to the oscillatory behavior of the energy
levels. Indeed, since it is the absolute value of the group
velocity that enters Eq.\ (\ref{heat-cond-Land-estim}), both
positive and negative slopes of the oscillating energy
$\varepsilon_\mu (k_z)$ as a function of $k_z$ in Eq.\
(\ref{spectrum-cylinder}) or in Fig.~\ref{fig:spectra} contribute
equally to the heat flux. This contribution is finite even for
very low temperatures and is determined by the number of crossings
of the constant energy line $\varepsilon \sim T$ by the
oscillating spectrum. Thus this number represents the number of
conducting modes $N_{eff}^{(R)}\sim N_{k_z}N_\mu $, where
$N_{k_z}$ is the number of crossings, as a function of $k_z$, of a
spectral branch with a given angular momentum $\mu$, and $N_\mu$
is the number of the angular momentum branches that cross this
line of energy $\varepsilon \sim T$. For large number of modes,
i.e., when the oscillation amplitude is larger than $\omega_0$ we
have\cite{PRL-2005} $N_{k_z}\sim k_F R (1-R/R_c) $ and $ N_\mu\sim
(2\Delta_0/\omega_0) e^{-2R/\xi} $. Finally, we get
\begin{equation}
N_{eff}^{(R)} \sim k_F R(1-R/R_c) (2\Delta_0/\omega_0) e^{-2R/\xi}
\ . \label{N_eff_R}
\end{equation}
If $R\sim\xi$, then $N_{eff}^{(R)}\sim N_{\rm Sh} \sim(k_F\xi)^2$
which coincides with the estimate of the number of transport modes
for a normal metal cylinder. We see therefore that the oscillations
of the vortex-core energy spectrum in a mesoscopic cylinder lead to
a drastic increase in the number of the conducting modes and,
accordingly, to an enhanced electron thermal conductance through the
vortex cores as compared to that in an infinite sample.

The temperature dependence of the effective number of modes
calculated using Eq.~(\ref{heat-cond-Land-estim}) is shown in
Fig.~\ref{fig:Neff-T-analyt}. The effective number of modes for
$T\gg \omega_0$ is
\begin{equation}
\label{Neff-T>omega0}
 N_{eff}^{(R)}(T)\simeq N^{\star} +\frac{27\zeta(3)}{2\pi^2}
\frac{T}{\omega_0} \ .
\end{equation}
Here $N^{\star}$ is constant as function of temperature; it is
small for $R>R_c$.  However $N^{\star}$ grows and approaches the
value $N_{eff}^{(R)}$ in Eq.\ (\ref{N_eff_R}) if the cylinder
radius decreases below $R_c$ and the amplitude of oscillation
increases. This asymptotic temperature dependence is observed
already for $T\gtrsim 0.5\, \omega _0$. The initial steep increase
in $N_{eff}$ in Fig.~\ref{fig:Neff-T-analyt} is associated with
the crossover into the regime of large number of modes when
temperature is increased up to $T\sim\omega_0$. The constant value
$N^{\star}$ leads to a linear temperature dependence of the heat
conductance $\kappa (T)$ in the range $T\gtrsim \omega_0$. The
linear in $T$ contribution can exceed the quadratic term given by
Eq. (\ref{Landauer-cond}) or by the second term in Eq.
(\ref{Neff-T>omega0}).

The effective number of modes  calculated using
Eq.~(\ref{heat-cond-Land-estim}) is shown in Fig.~\ref{fig:Neff-R}
as a function of the cylinder radius. Its non-monotonic dependence
on $R$ agrees reasonably well with our simple estimate in Eq.
(\ref{N_eff_R}).


\subsection{Exact quantum-mechanical analysis} \label{subsec-Exact}

A disadvantage of the above semiclassical picture is that it does
not provide a systematic way how to calculate the probability for
excitations to penetrate into the vortex core modes, which is
accompanied by reflection of excitations from the vortex ends at
the boundaries between normal reservoirs and the mesoscopic
superconductor. In this Section we derive the general
quantum-mechanical expression for the heat conductance in
superconductors directly from Eq.\ (\ref{encurrent/def}) and
demonstrate that it has the form of the Landauer formula with the
proper account of the scattering processes. Next we find the exact
solution of the BdG equations and calculate the vortex heat
conductance numerically.

\begin{figure}[hbt]
\centerline{\includegraphics[width=0.8\linewidth]{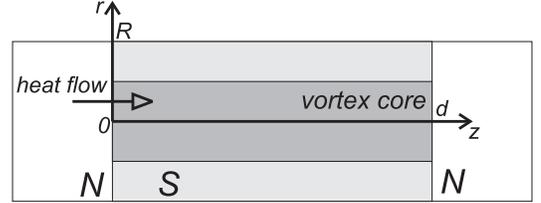}}
\caption{\label{fig:scheme} Heat transport through a vortex line.}
\end{figure}

We consider a  cylindrical superconducting slab of the thickness $d$
(Fig.~\ref{fig:scheme}) confined between the two normal leads
(reservoirs). The normal-superconducting interface is assumed
ideally transparent. The axis of a trapped vortex line coincides
with the superconducting cylinder axis. To calculate the heat
conductance we solve the scattering problem for electron-like ($U$)
and hole-like ($V$) quasiparticle wave functions within the BdG
theory.
It is convenient to consider four types of linearly independent
states:
(i) the state $(U_1,V_1)$ with an electron wave incident on a
superconductor from the left lead in the absence of other incident
electrons or holes;
(ii) the state $(U_2,V_2)$ with a hole wave incident on a
superconductor from the left lead;
(iii) the state $(U_3,V_3)$ with an electron wave incident on a
superconductor from the right lead; and
(iv) the state $(U_4,V_4)$ with a hole wave incident on a
superconductor from the right lead.

In the left lead, the first (electron) state has the form:
\begin{eqnarray}
U_1 &=& \frac{e^{i(\mu+1/2)\theta}}{\sqrt{2\pi}}g_{u,\mu
s}(r)e^{+ik_{z}z} \nonumber \\
&&+\frac{e^{i(\mu+1/2)\theta}}{\sqrt{2\pi}}
\sum_{s'}r_{uu,ss'}g_{u,\mu s'}(r) e^{-ik^\prime _{zu}z}\,
,\label{u-refl} \\
V_1 &=&\frac{e^{i(\mu-1/2)\theta}}{\sqrt{2\pi}}
\sum_{s'}r_{uv,ss'}g_{v,\mu s'}(r) e^{+ik^\prime _{zv}z}
\label{v-refl} \ .
\end{eqnarray}
Here $g_{u,\mu s}(r)$ and $g_{v,\mu s}(r)$ are the transverse
electron and hole modes of a normal metal wire with an angular
momentum $\mu$ and a radial quantum number $s$ respectively. The
transverse modes for each $\mu$ satisfy the boundary conditions
$g_{u,\mu s}(R) =g_{v,\mu s}(R) =0 $; they are normalized,
\[
\int_0^{R } g_{u,\mu s}g_{u,\mu s'}^*r \,dr = \int_0^{R } g_{v,\mu
s}g_{v,\mu s'}^*r \,dr = \delta_{s,s'}\ ,
\]
and have following form:
\begin{eqnarray}
g_{u,\mu s} &=&\frac{\sqrt{2} J_{|\mu+1/2|}\left(j_s^{(|\mu+1/2|)}
r/R\right)}{R J_{|\mu+1/2|+1}\left(j_s^{(|\mu+1/2|)}\right)} \, ;
\label{NM-modes} \\
g_{v,\mu s} &=&\frac{\sqrt{2} J_{|\mu-1/2|}\left(j_s^{(|\mu-1/2|)}
r/R \right)}{RJ_{|\mu-1/2|+1} \left( j_s^{(|\mu-1/2|)}\right)} \,.
\end{eqnarray}
Here $J_n$ is the Bessel function of $n$-th order, $j_s^{(n)}$ is
the $s$-th root of the equation ${J_n(j_s^{(n)})=0}$. The solution
Eqs.\ (\ref{u-refl}), (\ref{v-refl}) corresponds to the energy
\begin{equation}
\varepsilon(\mu,s,k_z) = \frac{\hbar^2}{2m}\left[ k_z^2+
\left(j_s^{(|\mu+1/2|)}/R\right)^{2} -k_F^2 \right]\ .
\end{equation}
The wave numbers of the scattered electron and hole waves are
$k^\prime _{z, uv} \equiv k^\prime _{z, uv}(\mu, s^\prime , k_z),$
where
\[
k^\prime_{z\, u,v}(\mu,s^\prime,k_z)=\sqrt{k_F^2
-\left(j_{s^\prime}^{(|\mu\pm 1/2|)}/R\right)^{2} \pm
2m\varepsilon/\hbar^2 } \ .
\]

The factors $r_{uu,ss'}$ and $r_{uv,ss'}$ are the matrices of
normal and Andreev reflection amplitudes for the processes with
large and small transfers of longitudinal momentum, respectively.
They are non-diagonal in $s$ and $s^\prime$ which accounts for
scattering with a small change in the transverse momentum caused
by an inhomogeneous vortex order parameter profile in the radial
direction. The change in the radial momentum at a given energy
$\varepsilon$ leads to a change in the longitudinal momentum such
that the scattered waves have momenta $k_{zu}$ and $k_{zv}$
slightly different from the momentum $k_z$ in the incident wave.

Similarly, the state with an incident hole wave is
\begin{eqnarray}
U_2 &=& \frac{e^{i(\mu+1/2)\theta}}{\sqrt{2\pi}}
\sum_{s'}r_{vu,ss'}g_{u,\mu s'}(r) e^{-ik^\prime _{zu}z}\,,
 \\
V_2 &=&\frac{e^{i(\mu-1/2)\theta}}{\sqrt{2\pi}} g_{v,\mu
s}(r)e^{-ik_{z}z} \nonumber\\
&&+\frac{e^{i(\mu-1/2)\theta}}{\sqrt{2\pi}}
\sum_{s'}r_{vv,ss'}g_{v,\mu s'}(r) e^{+ik^\prime _{zv}z}\ .
\label{v-ref2}
\end{eqnarray}
The energy of the incident hole is
\begin{equation}
\varepsilon(\mu,s,k_z) = -\frac{\hbar^2}{2m}\left[ k_z^2+
\left(j_s^{(|\mu-1/2|)}/R\right)^{2}-k_F^2 \right]\ .
\end{equation}
The transmitted state (iii) in the left lead is
\begin{eqnarray}
U_3 &=&\frac{e^{i(\mu+1/2)\theta}}{\sqrt{2\pi}}
\sum_{s'}t_{uu,ss'}g_{u,\mu s'}(r) e^{-ik^\prime _{zu}z}\,,
\nonumber \\
V_3 &=&\frac{e^{i(\mu-1/2)\theta}}{\sqrt{2\pi}}
\sum_{s'}t_{uv,ss'}g_{v,\mu s'}(r) e^{+ik^\prime _{zv}z}\, .
\label{u-transm}
\end{eqnarray}
The waves $U_4$, $V_4$ in the state (iv) have the same form with
the replacements $t_{uu,ss'}\rightarrow t_{vu,ss'}$ and
$t_{uv,ss'}\rightarrow t_{vv,ss'}$. The factors $t_{uu,ss'}$ and
$t_{vv,ss'}$ describe the particle-particle and hole-hole
transmission processes with a small change in the longitudinal
momentum while the factors $t_{uv,ss'}$ and $t_{vu,ss'}$ describe
the transmission with a large transfer of the longitudinal
momentum involving normal scattering processes.

BdG equations (\ref{BdG}) conserve the quasiparticle current
\[
{\bf J}_{qp}= \frac{1}{2m}\left[u^* (-i\hbar {\nabla}
-\frac{e}{c}{\bf A} )u
 -v^* (-i\hbar {\nabla} +\frac{e}{c}{\bf A})v +c.c.\right]  .
\]
For the incident and transmitted waves Eqs. (\ref{u-refl}),
(\ref{v-refl}), etc., this conservation law yields the sum rules
\begin{eqnarray}
&&\sum _{s^\prime} \left(k_{zu}^\prime |r_{uu,ss^\prime}|^2+
k_{zv}^\prime |r_{uv,ss^\prime}|^2 \right. \nonumber \\
&&\left. +k_{zu}^\prime|t_{uu,ss^\prime}|^2+
k_{zv}^\prime|t_{uv,ss^\prime}|^2\right) =k_{z} ,\quad \label{rule1} \\
&&\sum _{s^\prime} \left(k_{zv}^\prime |r_{vv,ss^\prime}|^2+
k_{zu}^\prime |r_{vu,ss^\prime}|^2 \right. \nonumber \\
&&\left. +k_{zv}^\prime |t_{vv,ss^\prime}|^2+k_{zu}^\prime
|t_{vu,ss^\prime}|^2\right) =k_{z} .\quad \label{rule2}
\end{eqnarray}
In the quasiclassical approximation, the transverse momentum is
conserved, therefore
\[
r_{uv,ss^\prime}= r_{uv,s }\delta_{s,s^\prime}\ , \;
r_{vu,ss^\prime}= r_{vu,s }\delta_{s,s^\prime} \ .
\]
At the same time, the processes with large longitudinal momentum
transfer are prohibited
\[
r_{uu,ss^\prime}=r_{vv,ss^\prime}=t_{uv,ss^\prime}=t_{vu,ss^\prime}=0
\ .
\]
However, in the exact BdG formalism used in our numerical
calculations, we keep the full set of probabilities including
$r_{uu,ss^\prime}$, $r_{vv,ss^\prime}$, $t_{uv,ss^\prime}$, and
$t_{vu,ss^\prime}$.

The heat flow through the cylinder cross section in the left lead
can be calculated summing up the states (i)--(iv):
\begin{widetext}
\begin{eqnarray}
I_{\cal E}&=&\sum_{\mu,s}\int\frac{dk_z}{2\pi} \int
r\,dr\,d\theta\,\frac{\hbar}{m}\left\{\frac{}{}\right.
\left[\varepsilon\cdot{\rm Im}\left\{U_1^*\nabla
U_1\right\}n_L(\varepsilon)- \varepsilon\cdot{\rm Im}\left\{V_1^*\nabla
V_1\right\}\cdot(1-n_L(-\varepsilon))
\right]\nonumber \\
&&+\left[\varepsilon\cdot{\rm Im}\left\{U_2^*\nabla
U_2\right\}n_L(\varepsilon)- \varepsilon\cdot{\rm Im}\left\{V_2^*\nabla
V_2\right\}\cdot(1-n_L(-\varepsilon))
\right] \nonumber\\
&&+\left[\varepsilon\cdot{\rm Im}\left\{U_3^*\nabla
U_3\right\}n_R(\varepsilon)- \varepsilon\cdot{\rm Im}\left\{V_3^*\nabla
V_3\right\}\cdot(1-n_R(-\varepsilon))
\right] \nonumber \\
&&+\left[\varepsilon\cdot{\rm Im}\left\{U_4^*\nabla
U_4\right\}n_R(\varepsilon)- \varepsilon\cdot{\rm Im}\left\{V_4^*\nabla
V_4\right\}\cdot(1-n_R(-\varepsilon)) \right]
\left.\frac{}{}\right\}\ .\label{heat-flow}
\end{eqnarray}
\end{widetext}
The sum is taken over the propagating quasiparticle states with
${\rm Im}\,k_z=0$ and $\varepsilon\geq 0$ (we thus take into account
both incident electrons and incident holes).

Using the sum rules Eqs. (\ref{rule1}), (\ref{rule2}) and
transforming to the energy integral, the heat current becomes
\begin{equation}
I_{\cal E}=\frac{1}{\pi\hbar}\int_0^\infty
\varepsilon\,[n_L({\varepsilon})-n_R(\varepsilon)]\,{\cal T}(\varepsilon)
\,d\varepsilon\,, \label{heatcurrent-ex}
\end{equation}
where the transfer factor
\begin{equation}
{\cal T}(\varepsilon)=\sum_{\mu, s,s'}\left[
\left|t_{uu,ss'}\right|^2+\left|t_{uv,ss'}\right|^2
+\left|t_{vu,ss'}\right|^2+\left|t_{vv,ss'}\right|^2 \right]
\label{Neff-S}
\end{equation}
is the sum of partial transmission probabilities. Finally, the
dimensionless heat conductance (the effective number of modes)
becomes
\begin{equation}
N_{eff}=\frac{\kappa}{\kappa_0} =-\frac{3}{\pi^2 T^2}\int_0^\infty
\varepsilon^2\frac{dn(\varepsilon)}{d\varepsilon} {\cal
T}(\varepsilon)\,d \varepsilon \,.\label{N-eff}
\end{equation}

Equations (\ref{heatcurrent-ex}), (\ref{Neff-S}), (\ref{N-eff})
are the exact quantum-mechanical expressions for the heat
conductance. They correspond to the Landauer
formula\cite{Butcher90} in which the transmission probabilities
are defined through the full scattering problem for excitations in
superconductors.

The semiclassical approach developed in the previous section
assumes that each particle or hole, incident on one of the
superconductor/normal-metal interfaces, with the energy and
momentum corresponding to the one or another propagating vortex
core mode is subsequently transmitted along this channel with a
probability $P_i$ to the second normal lead:
\[
\left|t_{uu,s }\right|^2=\sum_{s_i^+}P_i\delta_{s, s_i^+}
\]
provided the direction of the quasiparticle momentum along $z$
coincides with the direction of the group velocity in the vortex
core, and
\[
\left|t_{vv,s }\right|^2=\sum_{ s_i^-}P_i\delta_{s, s_i^-}
\]
in case where the direction of the quasiparticle momentum along
$z$ is opposite to the direction of the group velocity. The index
$i$ denotes the set of quantum numbers $\mu, k_z$; the sum is
taken over such quantum numbers $s_i^+ $ and $s_i^-$ which satisfy
the energy conservation equations $ \varepsilon _\mu
[k_z(s_i^\pm)]=\varepsilon $ for increasing or decreasing branches
of $\varepsilon _\mu $ as functions of $k_z$, respectively (we
emphasize that the quantum number $s$ is regarded as continuous in
the semiclassical approximation). Using $t_{uv}=t_{vu}=0$ one can
write Eq. (\ref{Neff-S}) as
\[
{\cal T}(\varepsilon) =\sum _\mu \int_0^{k_F}P_{\mu,k_z} \left|
\frac{\partial\varepsilon_\mu}{\partial k_z}\right|
\delta(\varepsilon-\varepsilon_\mu(k_z))\,dk_z\,.
\]
As a result, Eq. (\ref{heatcurrent-ex}) coincides with
Eq.~(\ref{heatcurrent-Land}). Under the assumption of complete
transmission $P_{\mu, k_z} =1$ the transfer factor ${\cal
T}(\varepsilon)$ is simply equal to the number of quasiparticle
modes propagating along the vortex core for a given energy
$\varepsilon$.

\subsection{Numerical results}

Here we present the results of numerical calculations of the
vortex heat conductance for a typical parameter
$\Delta_0/E_F=0.01$ ($k_F\xi=200$). The vortex core was
approximated by Eq.~(\ref{Core-model}) with $\xi_v=\xi$. The
critical radius estimated by Eq.~(\ref{Rc}) for the chosen set of
parameters is $R_c\simeq3.96\,\xi$. The calculations were carried
out for two values of $R=3.5\,\xi<R_c$ and $R=4.5\,\xi>R_c$, and
for several values of superconductor lengths $d/\xi$ varied from
16 up to 2048. The corresponding energy spectra for propagating
quasiparticle waves in a cylinder of an infinite length tabulated
using Eq.~(\ref{spectrum-model}) are shown in
Fig.~\ref{fig:spectra}.

\begin{figure}[t]
\centerline{\includegraphics[width=1\linewidth]
{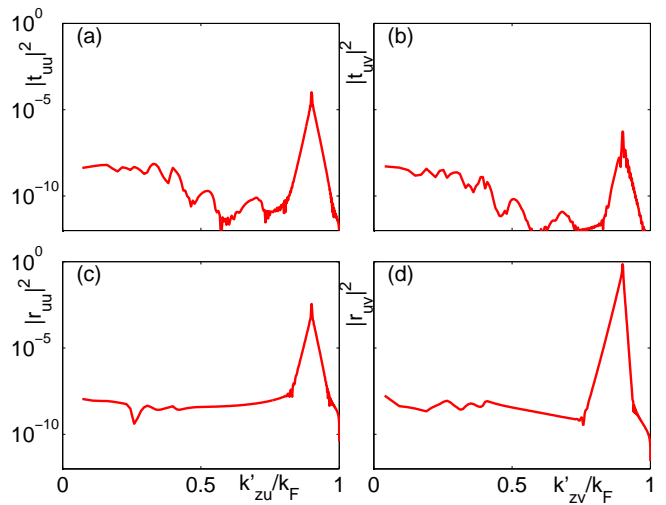}} \caption{\label{fig:T-matr} (a), (b) The
transmission $|t_{uu}|^2$, $|t_{uv}|^2$ and (c), (d) reflection
$|r_{uu}|^2$, $|r_{uv}|^2$
 probabilities, respectively, as functions of momenta $k^\prime_{zu}$ and
$k^\prime_{zv}$ of scattered waves for the incident electron-like
quasiparticle wave with $\mu=-1/2$, $k_z/k_F=0.9$, and
$\varepsilon =0.49\omega_0$. Parameters are: $R=4.5\xi$,
$d/\xi=16$.}
\end{figure}

\begin{figure}[t]
\centerline{\includegraphics[width=1.0\linewidth]{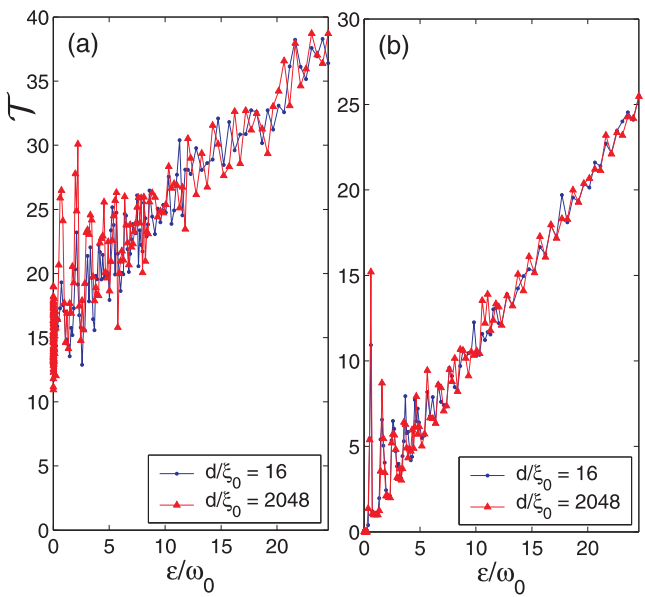}}
\caption{\label{fig:S2} (Color online) The transfer factor ${\cal
T}(\varepsilon)$ for different vortex line lengths $d$ (blue line
is for $d/\xi=16$ and red line is for $d/\xi=2048$) in the
cylinder with $R/\xi=3.5$ (a) and $R/\xi=4.5$ (b).}
\end{figure}

We calculate the scattering matrix of the normal modes incident on
the mesoscopic superconductor with a trapped vortex line using a
representation of the BdG operator in the truncated basis of the
normal-metal eigenfunctions. We solve the system of linear
equations which correspond to the boundary conditions at the
normal-metal/superconductor interfaces and to the conditions at
the infinity (radiation condition). Typical scattering matrix
coefficients as functions of the momentum of scattered wave are
plotted in Fig.~\ref{fig:T-matr}. The coefficients display peaks
around $k_{zu,v}=k_z$. The energy dependence of the transfer
factor ${\cal T}(\varepsilon)$ calculated using Eq.\
(\ref{Neff-S}) for two lengths $d$ of the vortex line is shown in
Figs.~\ref{fig:S2}. The linear in $\varepsilon/\omega_0$ term in
the transfer factor corresponds to the increasing number of CdGM
modes with $|\mu|\omega_0\leq\varepsilon$ which contribute to the
heat transport.

\begin{figure}[hbt]
\centerline{\includegraphics[width=1.0\linewidth]{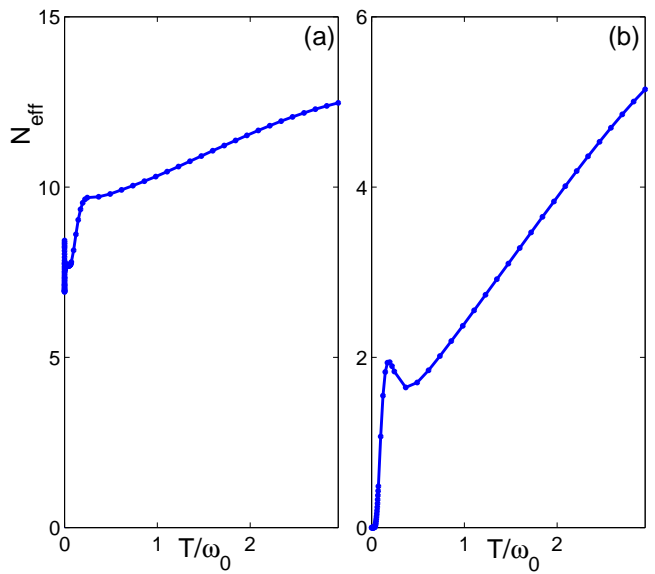}}
\caption{\label{fig:Neff-T} The effective number of modes
$N_{eff}(T)$ calculated numerically as a function of temperature
for the vortex line length $d/\xi_0=16$ in cylinders with
$R/\xi_0=3.5$ (a) and $R/\xi_0=4.5$ (b).}
\end{figure}

The temperature dependence of the heat conductance calculated
using Eq.\ (\ref{N-eff}) and plotted in Fig.~\ref{fig:Neff-T} is
qualitatively similar to that obtained on the basis of the
semiclassical model (see Fig.~\ref{fig:Neff-T-analyt}). In
particular, the slope of the linear asymptotic dependence
($T>\omega_0$) of $N_{eff}$ vs temperature is indeed well
described by the analytical approach. However, the value $N_{eff}$
calculated numerically is about seven times smaller than that
evaluated within the semiclassical approach with $P_{\mu,k_z}=1$.
Such deviation is obviously caused by a finite probability for
excitations to penetrate into the vortex core states at the
normal-lead/superconductor interfaces. The conductance peak at
$T^*\simeq0.2\omega_0$ which is most pronounced for the radius
$R>R_c$ (see Fig.~\ref{fig:Neff-T}b) is associated with the
contribution of the lowest oscillating CdGM level to the transfer
factor ${\cal T}(\varepsilon)$.

In our numerical calculations we do not observe the power-law
dependence of the thermal conductance on the thickness $d$
predicted in Ref.~[\onlinecite{KMV03}] which is associated with
the ballistic quasiparticle trajectories passing close to the
vortex axis. This power-law behavior is expected to dominate only
for much shorter $d$; the transport associated with such
trajectories appears to be closed already for our set of material
parameters and $d>16\xi$ since it would result in $N_{eff}\lesssim
10^{-3}$.

\section{Vortex in a cylinder with a rough
surface}\label{Sec-Roughness}

In the previous sections we calculated the electronic spectrum and
the heat transport for a sample with a perfect surface. One can
expect, however, that surface imperfections and roughness would
reduce the amplitude of spectrum oscillations. Making use of the
approach developed in Section \ref{Sec:spectrum} we analyze here
how surface roughness affects the vortex-core spectrum and the
heat transport mediated by the vortex core states. We show that
there exists a broad class of surface imperfections which do not
destroy mesoscopic spectrum oscillations completely; in this case
the heat transport through vortices still remains significantly
enhanced as compared to that in the bulk system.

Let us treat the potential $\hat V$ in Eq.
(\ref{eikonalequation2}) as a perturbation assuming the cylinder
radius to be sufficiently large. The solution to this equation
within the zero-order in $\hat V$ is a set of orthogonal and
normalized eigenfunctions Eq. (\ref{egenfunct-ideal}). The
first-order correction to the energy levels $\delta
\varepsilon\equiv \varepsilon-\epsilon (\mu) $ is
\begin{equation}
\delta\varepsilon =\frac{\Delta_0}{\pi\Lambda}
\int\limits_0^{2\pi} e^{-D_{\theta}}\cos\left[\alpha_\theta
-\pi\mu+\pi/2\right]d \theta \ .  \label{E-perturb}
\end{equation}

Consider several examples. Assume first that the cylinder has a
smooth surface $R(\theta)$. The main contribution to the integral
in Eq. (\ref{E-perturb}) comes then from the stationary phase
angles $\theta_j$ defined by the condition $\left. d\alpha _\theta
/d\theta \right|_{\theta_j}= 2k_\perp \left. d R/ d
\theta\right|_{\theta_j}=0 $. In the limit $k_\perp
R^{\prime\prime}\gg 1$ we thus obtain
\begin{eqnarray*}
\delta \varepsilon
&=&\frac{\Delta_0}{\Lambda}\sum_j\sqrt{\frac{2}{\pi k_\perp
|R^{\prime\prime}(\theta_j)|}}
e^{-D_{\theta_j}}\\
&&\times \cos\left[\alpha_{\theta_j}-\pi\mu+\pi/2+ \frac{\pi}{4}\,
{\rm sign}\, (R^{\prime\prime}(\theta_j))\right]  \ .
\end{eqnarray*}

For a simple realization $R(\theta)=R_0+a\cos(m\theta)$, where
$a\ll\xi$, the correction to the energy takes the form:
\[
\delta\varepsilon=\left(2\Delta_0/\Lambda\right)
e^{-D_0}\cos(\alpha_0-\pi\mu+\pi/2) J_0(2k_\perp a)
\]
where $\alpha_0=2k_\perp R_0$ and $D_0\equiv D(R_0)$. In the limit
$k_\perp a\gg 1$ we obtain
\begin{eqnarray*}
\delta\varepsilon &=&\left(2\Delta_0/\Lambda\right) (\pi k_\perp a
)^{-1/2}e^{-D_0}\\
&&\times \cos(\alpha_0-\pi\mu+\pi/2) \cos(2k_\perp a-\pi/4)
 \ .
\end{eqnarray*}
The amplitude of oscillations is decreased by a factor
$(k_Fa)^{-1/2} \ll 1$. Nevertheless, the amplitude $\delta
\varepsilon \gg \omega _0$ as long as $e^{-D_0} \gg
(k_Fa)^{1/2}(k_F\xi)^{-1}$. Note that the r.h.s. of this
inequality is a small parameter.

Let us now assume that the cylinder has a rough surface such that
the radius is $R(\theta)= R_0+ a(\theta)$, where $a(\theta)$ is a
random function of sample realizations with a zero ensemble
average $\langle a\rangle =0$ having a Gaussian distribution
\[
w_1(a)=\frac{1}{\sqrt{2\pi}\sigma}
\exp\left[-\frac{a^2}{2\sigma^2}\right] \ .
\]
Here $\sigma\ll \xi$ is the mean square value: $ \langle
a^2\rangle = \sigma^2 $. Averaging of Eq. (\ref{E-perturb}) gives
\begin{eqnarray*}
\langle \delta \varepsilon \rangle &=&\frac{\Delta_0}{\pi\Lambda}
e^{-D_0} \cos(\alpha_0-\pi\mu+\pi/2)\int _0^{2\pi}
\langle e^{2ik_\perp a(\theta)} \rangle d\theta \\
&=&\frac{2\Delta_0}{\Lambda} e^{-D_0} \cos(\alpha_0-\pi\mu+\pi/2)
 e^{-2k_\perp^2 \sigma^2} \ .
\end{eqnarray*}
This average vanishes for $k_F\sigma \gg 1$.

However, the energy level fluctuations are considerable. To prove
this we calculate the mean square energy fluctuation assuming a
Gaussian distribution for $a_1 = a(\theta_1)$ and $a_2 =
a(\theta_2)$:
\[
w_2(a_1,a_2)=\frac{1 }{2\pi\sigma^2\sqrt{1-K^2}}
\exp\left[-\frac{r^2}{2(1-K^2)\sigma^2}\right] \ ,
\]
where $ r=\sqrt{a_1^2+a_2^2-2Ka_1a_2} $ and
$K=K(\theta_1-\theta_2)=\langle a_1a_2\rangle/\sigma^2$ is the
correlation coefficient. This correlation coefficient should be an
even and $2\pi$-periodic function of angle: $K(\theta)
=K(-\theta)=K(\theta+2\pi)$. For $\theta_1=\theta_2$ we get $K=1$.
The correlation coefficient decays as we increase
$|\theta_1-\theta_2|$ value from $0$ to $\pi$.  We obtain
\begin{widetext}
\begin{eqnarray*}
\langle \delta \varepsilon^2 \rangle &\simeq &
\frac{\Delta_0^2}{\pi^2\Lambda^2} e^{-2D_0}
\int\limits_0^{2\pi}\int\limits_0^{2\pi}
\left<\cos(\alpha_{\theta_1}-\pi\mu+\pi/2)
\cos(\alpha_{\theta_2}-\pi\mu+\pi/2) \right> d \theta_1
 d \theta_2 \\
&=& \frac{\Delta_0^2}{2\pi^2\Lambda^2} e^{-2D_0}
\int\limits_0^{2\pi}\int\limits_0^{2\pi} \left[ \cos(2\alpha_0)
\left< e^{2ik_\perp (a_1+a_2)} \right> + \left< e^{2ik_\perp
(a_1-a_2)} \right> \right] d \theta_1
 d \theta_2 \ .
\end{eqnarray*}
\end{widetext}
For $k_\perp\sigma\gg 1$ the second term is only important. Using
the characteristic function for a Gaussian process
\[
\left< e^{i(u_1a_1+u_2a_2)} \right> =
e^{-\frac{1}{2}\sigma^2(u_1^2+u_2^2+2Ku_1u_2) }
\]
where $u_1=-u_2=2k_\perp$ we get
\[
\int\limits_0^{2\pi}\int\limits_0^{2\pi} \left< e^{2ik_\perp
(a_1-a_2)} \right> d \theta_1\, d \theta_2
=4\pi\int\limits_0^{\pi} e^{-4\sigma k_\perp^2(1-K(\theta))}\,
d\theta
\]
Expanding the correlation coefficient for small angles,
$K(\theta)\simeq 1-
|K^{\prime\prime}_{\theta\theta}(0)|\theta^2/2$, and introducing
the correlation radius $ \ell= R_0/\sqrt{|K^{\prime\prime}(0)|}$
we find
\begin{equation}
\langle \delta \varepsilon^2 \rangle \simeq \frac{\Delta_0^2
\ell}{\Lambda ^2 k_\perp\sigma R_0 (2\pi)^{1/2}} e^{-2D_0}
 \ .
\end{equation}
This expression is valid if the range of angles contributing to
the integral is small, $\theta \sim 1/\sigma k_\perp
\sqrt{|K^{\prime\prime}(0)|} \ll 1$. Taking $\ell\sim\sigma$ we
obtain
\begin{equation}
\sqrt{\langle \delta \varepsilon^2 \rangle} \simeq
\frac{\Delta_0}{\Lambda \sqrt{k_\perp R_0}  (2\pi)^{1/4}}
e^{-D_0}\gg |\langle \delta \varepsilon \rangle|
 \ .
\end{equation}
We see that the mean square fluctuation is much larger than the
CdGM interlevel spacing if the sample radius is not very large,
$e^{-D_0} \gg (k_FR_0)^{1/2}(k_F\xi)^{-1}$. Note again that the
r.h.s. here is a small parameter.

When the correlation radius increases and sample dimensions are
small $ R_0<\ell/k_\perp\sigma $ we can take $K\simeq 1$ for all
$\theta$. As a result
\begin{equation}
\sqrt{\langle \delta \varepsilon^2 \rangle} \simeq
(\Delta_0\sqrt{2}/\Lambda ) e^{-D_0}
 \ .
\end{equation}
In this limit the fluctuations of energy levels are universal in
the sense that they do not depend on $\ell$ and $\sigma$, though
do depend on the sample size.

To summarize, we see that imperfections on sample surface do not
always remove the spectrum oscillations. We have demonstrated that
the number of propagating quasiparticle modes responsible for the
heat transport in a mesoscopic sample can still be essentially
large for smooth surfaces with comparatively big variations in the
radius and also for Gaussian imperfections with fluctuations as
large as $\sigma\gg k_F^{-1}$.

\section{Conclusions}

We have investigated the electronic thermal transport along the
flux line traversing a mesoscopic superconducting sample having
the shape of a cylinder with the arbitrary cross section. We have
shown that the normal quasiparticle scattering at the sample
boundaries results in the essential increase in the number of
modes propagating along the vortex core. We have found that the
semiclassical version of the Landauer theory provides an adequate
qualitative description of the heat transport carried by these
modes. We have verified this conclusion by the detailed numerical
analysis of the exact quantum-mechanical scattering problem
described by the Bogoliubov--de Gennes theory. For small-radius
cylinders, such that $R\lesssim R_c$, the minigap in the vortex
spectrum is suppressed and the conductance becomes a linear
function of temperature in the range $T\gtrsim \omega _0$. This
linear term in the vortex conductance increases with the decrease
in the cylinder radius; it can strongly exceed the quadratic term
Eq.\ (\ref{Landauer-cond}) which usually dominates in the bulk
system at $\omega_0\ll T\ll T_c$.

We have analyzed how imperfections at the cylinder surface affect
the quasiparticle spectrum. We have found that there exists a wide
class of surface imperfections that give rise to only partial
reduction in the number of the conducting modes mediating the heat
conductance, even if the dimensions of imperfections exceed the
atomic scale $k_F^{-1}$.

\acknowledgments

This work was supported in part by the US DOE Office of Science
under contract No. W-31-109-ENG-38, by Russian Foundation for
Basic Research, the Program ``Quantum Macrophysics'' of the
Russian Academy of Sciences, Russian Science Support Foundation,
 and ``Dynasty'' Foundation.
 ASM acknowledges the support by the Academy of Finland.

\pagebreak

\end{document}